\documentclass[twocolumn,epjc3]{svjour3}          

\RequirePackage[T1]{fontenc}

\smartqed  

\RequirePackage{graphicx}
\RequirePackage{mathptmx}      
\RequirePackage{flushend}
\RequirePackage[numbers,sort&compress]{natbib}
\RequirePackage[colorlinks,citecolor=blue,urlcolor=blue,linkcolor=blue]{hyperref}

\journalname{Eur. Phys. J. C}

\begin{document}

\title{Anisotropic generalization of Buchdahl bound for specific stellar models}

\author{Ranjan Sharma \thanksref{e1,addr1} 
        \and 
        Arpita Ghosh \thanksref{e2,addr1}
        \and
        Soumik Bhattacharya \thanksref{e3,addr1}
				\and
				 Shyam Das \thanksref{e4,addr2}
}

\thankstext{e1}{e-mail: rsharma@associates.iucaa.in}
\thankstext{e2}{e-mail: arpitaghosh92727@gmail.com}
\thankstext{e3}{e-mail: soumik.astrophysics@gmail.com}
\thankstext{e4}{e-mail: dasshyam321@gmail.com}

\institute{Department of Physics, Cooch Behar Panchanan Barma University, Cooch Behar 736101, 
           West Bengal, India.\label{addr1}
          \and
          Department of Physics, Malda College, Malda 732101, West Bengal, India.\label{addr2}
   }

\date{Received: date / Accepted: date}

\maketitle

\begin{abstract}
Anisotropy is one factor that appears to be significantly important in the studies of relativistic compact stars. In this paper, we make a generalization of the Buchdahl limit by incorporating an anisotropic effect for a selected class of exact solutions describing anisotropic stellar objects. In the isotropic case of a homogeneous distribution, we regain the Buchdahl limit $2M/R \leq 8/9$. Our investigation shows a direct link between the maximum allowed compactness and pressure anisotropy vi-a-vis geometry of the associated $3$-space. 
\end{abstract}

\section{\label{sec1} Introduction}
What is the maximum permissible compactness of a self-gravitating compact stellar object? This question remains intriguing in the context of the stability of a stellar configuration, in particular. Buchdahl theorem \cite{buch59} addresses the question which states that no uniform density stars with radii smaller than $9/8M$ can exist. Otherwise, the central pressure diverges. In other words, for a stellar configuration in equilibrium, the Buchdahl bound implies $2M/R <8/9$. The interior of the star in this derivation is assumed to be a homogeneous distribution of perfect fluid whose exterior region is described by the Schwarzschild solution. Buchdahl bound prescribes an absolute constraint of the maximum allowed mass ($M$) to radius ($R$) ratio of a uniform density star as well as stellar configurations with radially decreasing energy density. Later on, by considering a variety of matter distributions in different background spacetimes, numerous studies have been made to determine the upper bound on the compactness of a star (for a ready reference, see \cite{dadh20} and references therein). Of particular interest is the recent work of Dadhich \cite{dadh20} where it has been shown that the Buchdahl compactness can be determined without any reference to the interior matter distribution. The upper limit in this calculation gets determined solely in reference to the Schwarzschild exterior solution for a neutral star and exterior Reissner-Nordstr\"om solution for a charged sphere. The probe also prescribes an upper bound on the charge to mass ratio $Q^2/M^2 \leq 9/8$. In a recent article, by developing a model for charged star which was shown to be a generalization of the uniform density Schwarzschild interior solution, Sharma {\em et al}\cite{sharma2021} regained the charged analogue of the Buchdahl compactness bound $$\frac{M}{R} = \frac{8/9}{(1+\sqrt{1-\frac{8\alpha^2}{9}})},$$ where $\alpha^2 = Q^2/M^2$. Obviously, the bound reduces to  $2M/R <8/9$ in the uncharged case. 

While numerous charged analogue of the Buchdahl bound is available in the literature (see, for example, \cite{giul08, mak01, boh07}), the study of the maximum mass-radius ratio for compact stars has been mostly confined to isotropic fluid spheres in which the tangential pressure ($p_{t}$) equals the radial pressure ($p_{r}$). However, theoretical investigations show that anisotropic stresses might exist beyond a certain density range at the interior of compact stars \cite{rude72,bow74,herr97}. In a recent article, it has been argued that pressure anisotropy cannot be ignored in the studies of relativistic compact stars as it is expected to develop quite naturally by the physical processes inside such ultra-compact stars \cite{Herrera2020}.  Earlier, Heintzmann and Hillebrandt \cite{hein75} claimed that no limiting mass could exist for a neutron star with an anisotropic matter distribution. However, considering an anisotropic distribution of matter, Guven and Murchadha \cite{guve99} determined an upper limit on $M/R$. The investigation showed that for a stellar configuration with monotonically decreasing density profile and having $p_{r} \geq p_{t}$, the upper bound $ \frac{M}{R} \leq \frac{8}{9}$ holds if the gravitational mass is replaced by the quasi-local mass. However, the bound can not be recovered for configurations with $p_{t}\geq p_{r}$. Mak {\em et al} \cite{mak02} developed an anisotropic stellar model by assuming the anisotropic stress to be proportional to $r^{2}$ and analyzed the subsequent bound.

The objective of the current investigation is to obtain an anisotropic generalization of the Buchdahl bound. We plan to obtain the bound for  particular class of solutions describing the interior of static spherically symmetric anisotropic stars. In our calculation, we choose two realistic stellar models developed by Maurya {\em et al} \cite{maur16} and Das {\em et al} \cite{das20}) so as to get an insight into the effects of anisotropy on the maximum bound of a star. Our choice of the particular class of solutions is motivated by the fact that both the solutions have their respective isotropic limits which enables one to regain the Buchdahl bound for a constant density star.

The paper is organized as follows: In sec.~\ref{sec2}, we outline the techniques adopted to generate the stellar models used in our analysis. In sec.~\ref{sec3}, we make use of the models to find anisotropic generalizations of the Buchdahl bound. In sec.~\ref{sec4}, concludes by highlighting the main results of our investigation. We also outline the scope of a further probe in this context.

\section{\label{sec2} Particular class of stellar solutions}
To develop the model of a static spherically symmetric star with anisotropic matter distribution, the line element describing the interior of the star is assumed to be of the form
\begin{equation}
ds^2 = -e^{\nu(r)}dt^2 + e^{\lambda(r)}dr^2 + r^2(d\theta^2 + \sin^2{\theta}d\phi^2),\label{eq1}
\end{equation}
where, $\nu(r)$ and $\lambda(r)$ are the unknown gravitational potentials. The energy-momentum tensor for the anisotropic fluid distribution is assumed in the form
\begin{equation}
T_{ab} = diag (-\rho,~p_{r},~p_{t},~p_{t}),\label{eq1a}
\end{equation}
where $\rho$, $p_{r}$ and $p_{t}$ are the energy density, radial and tangential pressure, respectively. The comoving fluid velocity is given by $\displaystyle u^{i} = e^{-\nu /2} \delta^i _{0}$. The Einstein field equations for the line element (\ref{eq1}) and the energy-momentum tensor (\ref{eq1a}) are then obtained as
\begin{eqnarray}
\rho &=& \frac{1}{r^2} (1-e^{-\lambda}) + \frac{\lambda'}{r} e^{-\lambda}, \label{eq1b}\\
p_r &=& - \frac{1}{r^2} (1-e^{-\lambda}) + \frac{\nu'}{r} e^{-\lambda}, \label{eq1c}\\
p_t &=& \frac{e^{-\lambda}}{4} \Big(2 \nu'' + {\nu'} ^2 -\nu' \lambda' + \frac{2 \nu'}{r} 
	- \frac{2 \lambda'}{r} \Big),\label{eq1d}
\end{eqnarray}
where a $(')$ denotes differentiation with respect to $r$. $\Delta = p_t - p_r$ denotes anisotropy in this construction. To close the system of equations (\ref{eq1b}) - (\ref{eq1d}), following techniques are adopted:

\subsection{\label{sub1} \textbf{Solution developed by Das {\em et al} \cite{das20}}}
To solve the system, one introduces the Vaidya and Tikekar \cite{vaid82} metric ansatz
\begin{equation}
e^\lambda = \frac{1-K(\frac{r^2}{L^2})}{1 -\frac{r^2}{L^2}},\label{eq2}
\end{equation}
where $K$ and $L$ are constant curvature parameters. The Vaidya and Tikekar metric ansatz is motivated by the observation that the geometry of the $t=$ constant hypersurface of the associated spacetime when embedded in a $4$-dimensional Euclidean space is not spherical but spheroidal in nature. The parameter $K$ in (\ref{eq2}) denotes the departure from the sphericity of associated $3$-space. The anstaz has found huge application for the development and studies of relativistic stellar models like neutron stars. Earlier, making use of the  Vaidya and Tikekar metric ansatz, Sharma {\em et al} \cite{shar06} obtained the maximum permissible mass and radius of a relativistic compact star for a given surface density. An upper bound on mass to radius ratio for an anisotropic stellar model was obtained by Maurya {\em et al} \cite{maur19} where the Buchdahl-Vaidya-Tikekar metric ansatz was utilized. 

Since anisotropy provides an additional degrees of freedom in this construction, in addition to the Vaidya and Tikekar ansatz, Das {\em et al} \cite{das20} also utilized the Karmarkar's embedding condition of class-I \cite{kar48} to determine the unknown metric potential $e^{\nu(r)}$. The procedure yields 
\begin{equation}
e^\nu =\Big[C+D \sqrt{(K-1)(r^2-L^2)}\Big]^2.\label{eq3}
\end{equation}
where $C$ and $D$ are constants to be determined from the boundary conditions. Subsequently, physical quantities are obtained as
\begin{eqnarray}
\rho &=& \frac{(K-1)(K r ^2-3L^2)}{(L^2-K r^2)^2},\label{eq4}\\
p_r &=& \frac{(K-1)\Big[D(K-3)\sqrt{r^2-L^2}+C\sqrt{K-1}\Big]}{(L^2-K r^2)\Big[D(K-1)\sqrt{r^2-L^2}+C\sqrt{K-1}\Big]},\label{eq5}\\
p_t &=& \frac{(K-1)}{(L^2-K r^2)^2\Big[D(K-1)\sqrt{r^2-L^2}+C\sqrt{K-1}\Big]}
\nonumber \\ 
&+& \Big[D(K-3)L^2\sqrt{r^2-L^2}+D K r^2\sqrt{r^2-L^2} \nonumber\\
&+& C L^2\sqrt{K-1}\Big],
\label{eq6}\\
\Delta &=& \frac{(K-1)K r^2\Big[D(K-2)\sqrt{r^2-L^2}+C\sqrt{K-1}\Big]}{(L^2-K r^2)^2\Big[D(K-1)\sqrt{r^2-L^2}+C\sqrt{K-1}\Big]}.\label{eq7}
\end{eqnarray}
The mass contained within a radius $r$ is obtained as
\begin{equation}
m(r) = \frac{(K-1)r^3}{2(L^2-K r^2)}.\label{eq8}
\end{equation}
Eq.~(\ref{eq7}) shows that that while $K$ denotes deviation from sphericity of the associated $3$-space, the parameter also turns out to be a measure of anisotropy in this formulation. Most importantly, the $K=0$ case suggests spherical homogeneous distribution with zero anisotropy and in that case the solution reduces to
\begin{eqnarray}
ds^2 &=& -\Big[C - DL \sqrt{1-\frac{r^2}{L^2}}\Big]^2 dt^2 + \frac{1}{1-\frac{r^2}{L^2}} dr^2 \nonumber\\
&& r^2(d\theta^2 + \sin^2{\theta}d\phi^2), \label{eq8a}
\end{eqnarray}
which is the Schwarzschild interior solution for an incompressible fluid.

For a given $K$, the constants of the model, namely $C$, $D$ and $L$ do get fixed by (i) matching the interior solution to the Schwarzschild exterior metric across the boundary  ($R$) and $(ii)$ by imposing the condition that the radial pressure should vanish at a finite boundary i.e., $p_r(r=R) = 0$.  In terms of total mass $M$ and radius $R$, the constants are obtained as
\begin{eqnarray}
L&=& R \sqrt{\frac{R+2 K M-K R}{2 M}},\label{eq9}\\
C &=&(R-2M)(3 - K)\sqrt{\frac{R (1 - K)}{8 M(L^2-R^2)}},\label{eq10}\\
D &=&-\sqrt{\frac{M}{2 R^3}}.\label{eq11}
\end{eqnarray}

\subsection{\label{sub2} \textbf{Solution developed by Maurya {\em et al} \cite{maur19}}}
Maurya {\em et al} \cite{maur16} assumed a specific form of the metric potential $g_{rr}$ 
\begin{equation}
e^\lambda = 1+ \frac{(a-b)r^2}{1+br^2}, \label{eq12}
\end{equation}
where $a$ and $b$ are constants with $a \neq b$. Using (\ref{eq12}) and Karmarkar's embedding condition, it is possible to find a closed form solution for the system (\ref{eq1b}) - (\ref{eq1d}) in the form
\begin{equation}
e^\nu = \Big[\frac {Ab + B \sqrt{(a-b)} \sqrt{(1+br^2)}}{b}\Big]^2,\label{eq13}
\end{equation}
where $A$ and $B$ are constants to be determined from the boundary conditions. Subsequently, one obtains
\begin{eqnarray}
\rho &=& (a-b) \Big[\frac{(3+ar^2)}{(1+ar^2)^2}\Big],\label{eq14} \\
p_r &=& (a-b) \Big[\frac{-Ab +B(3b-a)\sqrt{\frac{(1+br^2)}{(a-b)}}}
	{(1+ar^2)[Ab + B \sqrt{(a-b)(1+br^2)}]}\Big], \label{eq15} \\
p_t &=& \frac{\sqrt{(a-b)}}{(1+ar^2)^2[Ab + B \sqrt{(a-b)(1+br^2)}]} 
\nonumber\\
&& \Big[-Ab \sqrt{(a-b)} +\frac{B}{\sqrt{(1+br^2)}}(3b-a+a b^2 r^4)\Big].\label{eq16}
\end{eqnarray}
The anisotropy factor is obtained in the form
\begin{equation}
\Delta = ar^2 \sqrt{(a-b)} \Big[\frac{Ab \sqrt{(a-b)} +\frac{B}{\sqrt{(1+br^2)}}(a-2b)}
{(1+ar^2)^2 [Ab + B \sqrt{(a-b)(1+br^2)}]}\Big],\label{eq17}
\end{equation}
which shows that the anisotropy vanishes for $a=0$.

The boundary conditions mentioned above determine the unknown constants as
\begin{eqnarray}
A &=& \frac{(3b-a)}{2b} \sqrt{\Big[\frac{1+bR^2}{1+aR^2}\Big]},\label{eq18}\\
B &=& \frac{1}{2} \sqrt{\frac{a-b}{1+aR^2}}. \label{eq19}
\end{eqnarray}

The total mass $M$ within a radius $R$ is obtained as
\begin{equation}
M = \frac{R}{2} \Big[1 - \frac{1+bR^2}{1+aR^2}\Big]. \label{eq20}
\end{equation}
Note that the solution contains four constants $a,~b,~A,~B$. For given values of $M$ and $R$, three of the constants do get fixed while one parameter remains free. In our analysis, we assume the free parameter to be $a$. It is noteworthy that $b \neq 0$ in this model as $A$ becomes infinite for $b=0$. For $a =0$, equation (\ref{eq20}) suggests that we must have $b < 0$. Most importantly, by setting $a=0$ in Eq.~(\ref{eq12}) and setting $b = - \frac{1}{L^2}$, we regain the Schwarzschild incompressible fluid solution. 

Thus, we have two sets of anisotropic stellar solutions, both having the Schwarzschild interior solution as an isotropic limiting case. It can be shown that both the solutions satisfy the energy conditions and other physical requirements of a realistic star. In the following section, making use of the anisotropic solutions, we intend to obtain an anisotropic generalization of the Buchdahl bound.  
 
\section{\label{sec3} Anisotropic generalization of Buchdahl bound}
It is noteworthy that the solutions to be utilized to find a new bound should fulfil the physical requirements that density and pressure should be non-negative ($\rho,~p \geq 0$) and density should be a decreasing function of radial distance ($\frac{d\rho}{dr} > 0$). The generalized bound in the presence of anisotropic stress is deduced by demanding that the pressure must not diverge at the centre (note that at $r=0$, $p_r=p_t$). In the following sub-sections, we derive two separate bounds on the mass to radius ratio.  

\subsection{\label{sub3} \textbf{Generalized Buchdahl bound utilizing the solution obtained in (\ref{sub1})}}
\begin{figure} 
\centering
\includegraphics [width=.5\textwidth]{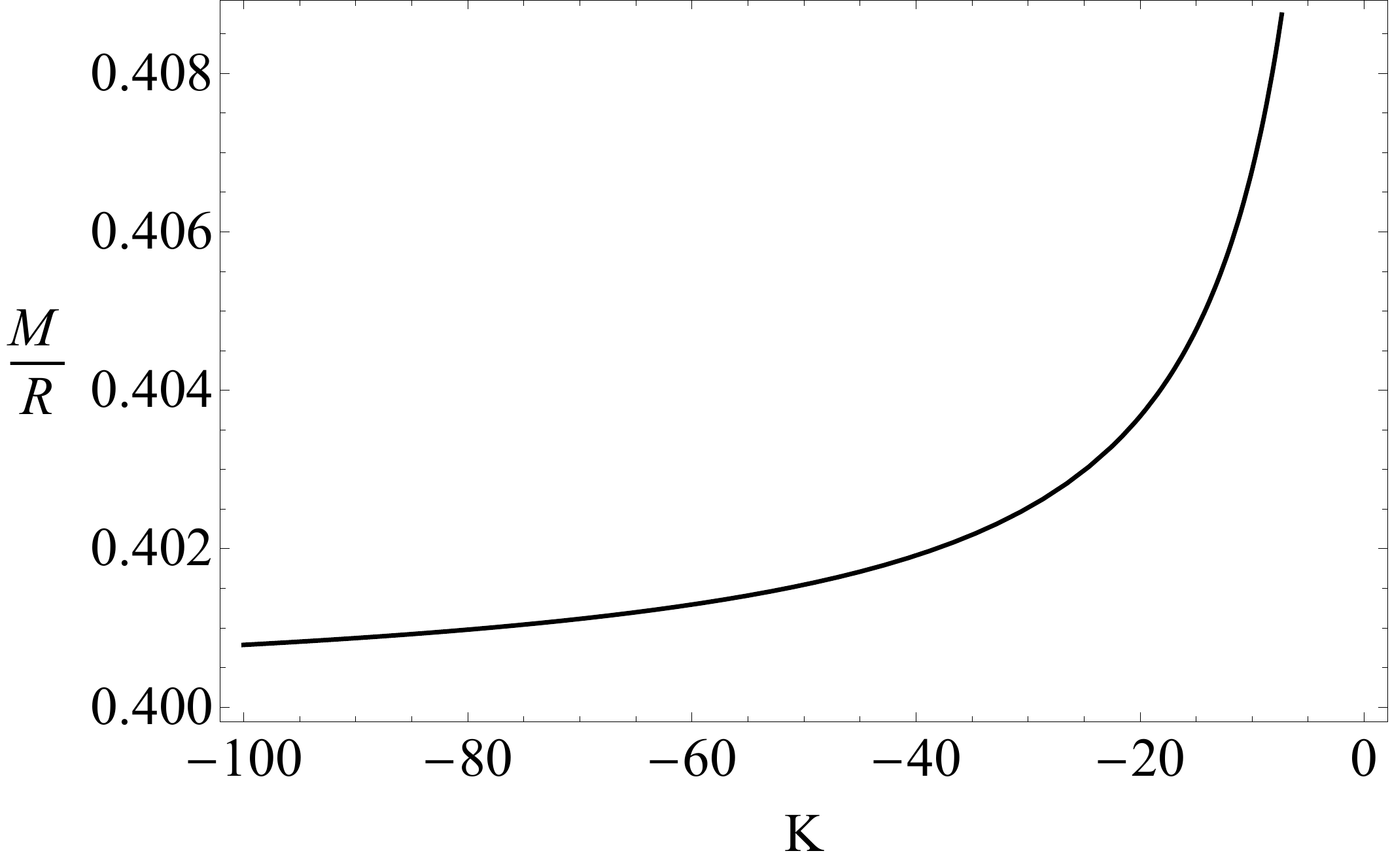}
\caption{Upper bound on $M/R$ plotted against the anisotropic parameter $K$.}
\label{f1}
\end{figure}

In eq.~(\ref{eq5}), the condition that central pressure must not diverge implies that we must have
\begin{equation}
\Big[C\sqrt{1-K}-D L(1-K)\Big] \geq 0. ~~~~~~~(L\neq 0) \label{eq21}
\end{equation}
To find the bound on $M/R$, we substitute the values of $C$,~$D$ and $L$ in eq.~(\ref{eq21}) and consider the extreme case (i.e., equate the left hand side to zero) which yields
\begin{equation}
\frac{(R-2M)(3+K^2-4K)}{(K-1)\sqrt{(R+2 K M-K R)(R+2 K M-K R-2M)}} = 1.\label{eq23}
\end{equation}
We rewrite (\ref{eq23}) as 
\begin{equation}
(9-5K) y^2+(9 K-17) y+(8-4 K) = 0,\label{eq24}
\end{equation}
where $\displaystyle y =\frac{2 M}{R}$. Solution of eq.~(\ref{eq24}) eventually leads to the following inequality
\begin{equation}
y = \frac{2 M}{R} \leq \frac{4(K-2)}{(5 K-9)}.\label{eq25}
\end{equation}
Since the parameter $K$ can be linked to anisotropy in this model, eq.~(\ref{eq25}) can be treated as a generalization of the Buchdahl bound for an anisotropic stellar configuration. In Fig.~(\ref{f1}), the bound on $M/R$ against $K$ is shown. The plot indicates that as anisotropy increases, the upper bound decreases. Since $K$ is also a measure of departure from homogeneous isotropic spherical distribution, the result also shows that a departure from spherical geometry reduces the upper bound on compactness. Setting $K = 0$, we regain the bound $2M/R \leq 8/9$.     

\subsection{\label{sub4} \textbf{Generalized Buchdahl bound utilizing the solution obtained in (\ref{sub2}):}}
\begin{figure} 
\centering
\includegraphics [width=0.5\textwidth]{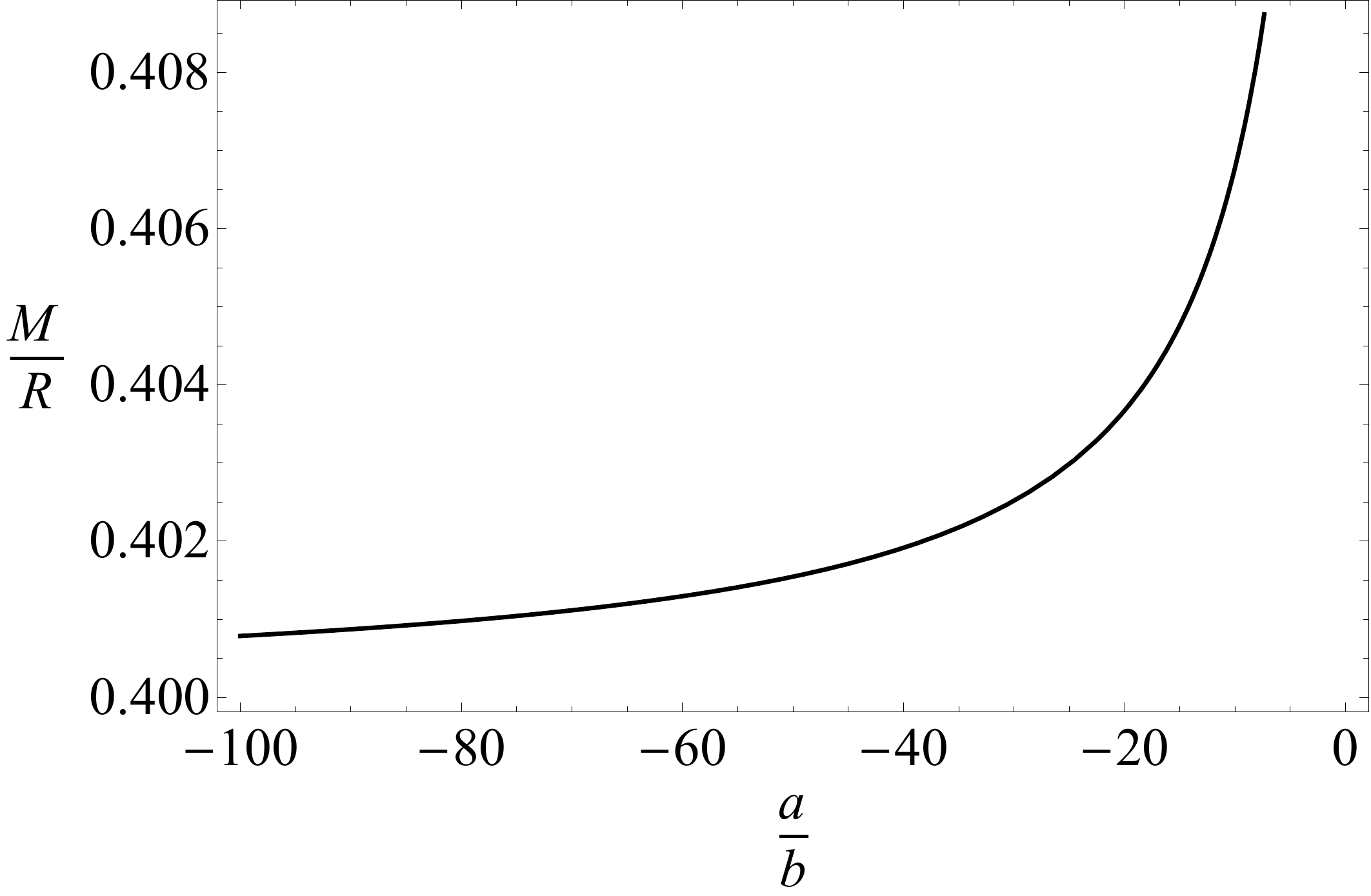}
\caption{Upper bound on $M/R$ plotted against the anisotropic parameter $a/b$.}
\label{f2}
\end{figure}

We utilize the solution obtained by Maurya {\em et al} and follow the same technique to make an anisotropic generalization of the Buchdahl bound. In this case, applying the condition that central pressure must not diverge in eq.~(\ref{eq15}), we obtain the following constraint
\begin{equation}
Ab + B \sqrt{(a-b)} \geq  0. \label{eq27}
\end{equation}
We substitute the values of $A$ and $B$ in (\ref{eq27}) and in the extreme case obtain the following relation
\begin{equation}
1 - \frac{2M}{R} = \Big(\frac {a-b}{3b-a}\Big)^2 \frac{1}{1+aR^2}, \label{eq28}
\end{equation}
which eventually leads to the following bound 
\begin {equation}
\frac {2M}{R} \leq \frac{4(\frac{a}{b})^2-12(\frac{a}{b})+8}{5(\frac{a}{b})^2-14(\frac{a}{b})+9}, \label{eq29}
\end {equation}
where we have substituted
\begin{equation}
R^2 =  \frac {4a-8b}{(3b-a)^2}. \label{eq30}
\end{equation}
Eq.~(\ref{eq29}) is an anisotropic generalization of the Buchdahl bound. Setting $a=0$ in eq.~(\ref{eq29}), we  regain the bound $2M/R \leq 8/9$. In Fig.~(\ref{f2}), the variation of the maximum bound on compactness with respect to the dimensionless anisotropic factor $a/b$ is shown. It is interesting to note that the maximum bound decreases with the increase of anisotropy. In fact, both the models, exhibit identical behaviour.
 
\section{\label{sec4} Discussion:}
The key results of our investigation are outlined below:
\begin{itemize}
\item Making use of particular anisotropic stellar models, we have obtained anisotropic generalizations of the Buchdahl bound. In the isotropic case, both the solutions correspond to the Schwarzschild incompressible fluid solution, and consequently, the Buchdahl bound $M/R \leq 4/9$ is regained once the anisotropy is switched off. For any physically reasonable stellar configuration, the generalized bound is obtained by demanding that the central pressure must not be infinite.

\item Even though we have considered two different class of solutions, the impact of anisotropy is similar in both cases. Anisotropy reduces the upper bound on compactness, and it never exceeds the value of $2M/R =8/9$.  
\end{itemize}

It remains to be seen whether the effects of anisotropy remain identical in all kinds of anisotropic stellar models. It should be stressed here that we have considered models where $p_t > p_r$. It remains to be seen whether a similar outcome would follow for stellar configurations having $p_r > p_t$. It is also a matter of further probe whether the technique adopted in this paper provides the most stringent bound on compactness. All these issues are beyond the scope of the current investigation and will be taken up elsewhere.

\begin{acknowledgements}
RS and SD gratefully acknowledge support from the Inter-University Centre for Astronomy and Astrophysics (IUCAA), Pune, India, under its Visiting Research Associateship Programme.
\end{acknowledgements}

\section{Data repository}
The data underlying this article is available in the public domain as cited in the references.


\begin{thebibliography}{99}

\bibitem{buch59}  H. A. Buchdahl, \textit{Phys. Rev.}, \textbf{116}, 1027 (1959)

\bibitem{dadh20} N. Dadhich, \textit{JCAP}, \textbf{4} 035 (2020)

\bibitem{sharma2021} R. Sharma, N. Dadhich, S. Das and S. D. Maharaj, {\it Eur. Phys. J C} {\bf81}, 79 (2021)

\bibitem{giul08} A. Giuliani, T. Rothman, \textit{Gen. Relativ. Grav.}, \textbf{40}, 1427 (2008)

\bibitem{mak01} M. K. Mak, P. N. Dobson and T. Harko, \textit{Euro. Phys. Lett.}, \textbf{56}, 762 (2001)

\bibitem{boh07} C. G. B\"ohmer and T. Harko, \textit{Gen. Relativ. Grav.}, \textbf{39}, 757 (2007)

\bibitem{rude72}  R. Ruderman, \textit{Ann. Rev. Astron. Astrophys.}, \textbf{10}, 427 (1972)

\bibitem{bow74} R. L. Bowers and E. P. T. Liang, \textit{Astrophys. J.}, \textbf{188}, 657 (1974)

\bibitem{herr97} L. Herrera and N. O. Santos, \textit{Phys. Rep.}, \textbf{286}, 53 (1997)

\bibitem{Herrera2020} L. Herrera, {\textit Phys. Rev. D}, \textbf{101}, 104024 (2020)

\bibitem{hein75} H. Heintzmann and W. Hillebrandt, \textit{Astron. Astrophys.}, \textbf{38}, 51 (1975)

\bibitem{guve99} J. Guven and N. O. Murchadha, \textit{Phys. Rev. D}, \textbf{60}, 084020 (1999)

\bibitem{mak02} M. K. Mak, P. N. Dobson, T. Harko and L. Z. Fang, \textit{Int. J. Mod. Phys. D}, \textbf{11}, 207 (2002)

\bibitem{das20} S. Das, R. Sharma, K. Chakraborty and L. Baskey, \textit{Gen. Rel. Grav.}, \textbf{52}, 101 (2020)

\bibitem{maur16} S. K. Maurya, Y. K. Gupta, T. T. Smita and F. Rahaman, \textit{Eur. Phys. J. A}, \textbf{52}, 191 (2016)

\bibitem{vaid82} P. C. Vaidya and R. Tikekar, \textit{J. Astrophys. Astro.}, \textbf{3}, 325 (1982)

\bibitem{shar06} R. Sharma, S. Karmakar and S. Mukherjee, \textit{Int. J. Mod. Phys. D}, \textbf{15}, 405 (2006)

\bibitem{maur19} S. K. Maurya, S. D. Maharaj, J. Kumar, A. K. Prasad, \textit{Gen. Relativ. Gravit.}, \textbf{51}, 86 (2019)

\bibitem{kar48} K. R. Karmarkar, \textit{Proc. Ind. Acad. Sci. A}, \textbf{27}, 56 (1948)

\end{thebibliography}
\end{document}